# Phonon thermal transport outside of local equilibrium in nanowires via molecular dynamics

Ya Zhou and Alejandro Strachan

School of Materials Engineering, Purdue University, West Lafayette, Indiana 47907, USA

Abstract

We study thermal transport through Pt nanowires that bridge planar contacts as a function of wire length and vibrational frequency of the contacts. When phonons in the contacts have lower average frequencies than those in the wires thermal transport occurs under conditions away from local equilibrium with low-frequency phonons experiencing a higher thermal gradient than high-frequency ones. This results in a size-dependent increase in the effective thermal conductivity of the wire with decreasing vibrational frequencies of the contacts. The interfacial resistivity when heat flows from the wire to the contact is also size-dependent and has the same physical origin in the lack of full equilibration in short nanowires. We develop a model based on a 1D atomic chain that captures the salient physics of the MD results.

**I. INTRODUCTION**

Low dimensional materials such as superlattices and nanowires exhibit transport properties not achievable in the bulk making them attractive for a variety of applications including energy conversion and optoelectronics [1-9]. For example, efficient thermoelectric devices require separate control of electrical and thermal transport since the figure of merit is proportional to the ratio of electrical to thermal conductivities. For such applications, nanoengineering has resulted in improved performance, see for example Refs. [2,4,10], with some of the most impressive recent advances originating from a reduction in phonon thermal conductivity. Nano-scale superlattices [6,7,11,12] and wires [1,2,5,13], where phonon thermal conduction is dominated by interfaces and transport between dissimilar materials, are among today's most promising nanostructures [12]. Thus, a physics-based understanding of thermal transport in nanostructured materials is highly desirable. Furthermore, many applications require the integration of materials with very different vibrational properties and transport across their interfaces exhibits interesting physics when size of the components is in the nanoscale [14].

While it is widely accepted that sub-continuum processes can lead to transport away from local-equilibrium in nanostructured materials, the nature of this non-equilibrium state is not well understood. Such lack of local equilibrium can manifest itself, for example, in thermal resistivity being dependent on the details of how energy is transferred to a nanoscale material [14]. While significant progress has been made in this field in recent years several questions remain unanswered; for example how size affects the non-equilibrium thermal transport. In this paper we use non-equilibrium molecular dynamics (NEMD) simulations to characterize phonon thermal transport through Pt nanowires sandwiched between parallel contacts. We focus on the role of the vibrational properties of the contacts and nanowire length, and characterize the nature of the non-equilibrium transport inside the wires. In order to vary the vibrational properties of the contacts without introducing additional changes at the interfaces (such as defects), we use a model system where the contacts are composed of an artificial material, denoted Pt*, that differs from Pt only in its atomic mass. For low vibrational frequency contacts, low-frequency modes in the nanowire experience a larger temperature gradient than high-frequency ones near the contacts. This results in noticeable, size-dependent changes in the wires' effective thermal conductivity and size-dependent interfacial resistance.

## II. SIMULATION DETAILS

### A. Nanowire-contact model systems

We perform thermal transport simulations using NEMD [15] for a series of Pt nanowires sandwiched between two Pt* contacts. Pt* is identical to Pt except for its atomic mass ($M^*$) that we vary in the range of $0.1M$ to $10M$; $M$ being the atomic mass of Pt. A representative configuration is shown in Figure 1. The nanowires are oriented along the [112] crystallographic direction, and have a diameter $d$=2.3 $nm$ and lengths between 2.4 and 71.7 $nm$. The wires bridge two Pt* contacts of 3.8 $nm$ thickness and $4.7 \times 4.7$ $nm^2$ cross-section; all interfaces are epitaxial, i.e. defect free. Periodic boundary conditions are imposed in the plane perpendicular to the wire axes and the systems are finite in the direction parallel to the wire axes. These structures mimic an experiment/device setup consisting of a nanowire array between two parallel contacts.

Interactions between Pt atoms are described with a many body potential of the embedded atom model type [16].

**B. Non-equilibrium MD simulations of thermal transport**

Thermal transport is simulated by imposing a known heat flux between the contacts and conductivity is computed from this flux and the local temperature profile after steady state is achieved [15]. Interfacial resistance between the wire and the contacts introduces temperature jumps across the interfaces and two temperature differences can be defined: i) the difference between the hot and cold contacts; and ii) the temperature gradient within the nanowire. We can thus compute two effective phonon thermal conductivities from Fourier's law: the overall thermal conductivity of the nanostructure and that of the wire itself [14].

The effective nanowire temperature profile is computed by spatial decomposition of the wire along its axis in 0.48 $nm$ thick sections. The average kinetic energy $\langle K_i \rangle$ in section $i$ is computed and the corresponding effective temperature is obtained as $3/2NkT_i^{eff} = \langle K_i \rangle$. This effective local temperature can only be called a true local temperature when local equilibrium applies. The effective temperature gradient in the wire is obtained from a linear fit of the $T_i^{eff}$ neglecting the two sections adjacent to each contact.

Previous work on the role of interfaces in thermal transport focused on phonon scattering and transmission [12,17]; our goal is to characterize phonon populations during transport. It is important to highlight that the NEMD approach we use makes no approximation regarding phonon scattering mechanisms, local equilibrium, or anharmonicities in atomic interactions due to finite temperatures.

## II. Thermal conductivity of nanowires: role of size and contact properties

### A. Non-equilibrium MD results

Interfacial scattering of phonons is enhanced as the nanowire-contact atomic mass difference increases and, thus, the computed overall thermal conductivity exhibits a maximum at $M^*/M=1$ and decreases monotonically as $M^*/M$ deviates from 1. The effective thermal conductivity of the nanowires in the presence of the contacts reveals more complex physics; Figure 2 shows its dependence on reduced mass of the contact for various wire lengths. As expected thermal conductivity of wires increases with wire length due to reduced boundary scattering in the axial direction [5,12]. More interestingly, for a given length the effective thermal conductivity of nanowires increases with the atomic mass of the contacts (even though the nanowires are identical). This effect is more pronounced for short wires. For example, thermal conductivity of a 4.8 *nm* long nanowire increases by ~ 60% as $M^*/M$ increases from 1 to 10. The shortest wire studied (2.4 *nm* long) deviates from this general trend due to the numerical uncertainties associated with the calculation of temperature gradient on short distances [14].

### B. Frequency-dependent temperature gradients during transport

To understand the physical origin of the thermal conductivity enhancement with increasing atomic mass of the contacts, we compute the vibrational power spectra (the kinetic energy in modes of frequency $\omega$ as a function of $\omega$) of nanowires and contacts using the velocity spectrum [18]. The inset of Fig. 3 (c) shows the equilibrium power spectra for a $M^*/M=10$ system where we separate the contributions of the contact and wire atoms; the heavy mass of the contacts leads to a red shift in the spectrum. To characterize how the spectra of incoming and outgoing phonons during thermal transport affect the phonon populations inside the wire we compute the power spectrum under equilibrium and thermal transport conditions; spatial resolution during transport is achieved using the same binning described above. We then divide the nanowire power spectrum into three regimes of low, intermediate, and high frequencies, see inset of Fig. 3 (c), and compute the kinetic energy content in each frequency range. Assuming that the phonon density of states (DoS) of the wire is not affected by thermal transport, at position $z$ along wire axis, the local nanowire temperature associated with modes in each of the frequency ranges ($\Delta\omega$) can be computed as $T_{trans}(\Delta\omega, z) = \frac{S_{trans}(\Delta\omega, z)}{S_{eq}(\Delta\omega, z)} 300K$ [14], where $S$ is the corresponding kinetic energy content and 300 $K$ is the temperature of all modes in equilibrium, the z-dependence of the equilibrium spectrum originates from interactions with the contacts.

Figure 3 shows local temperature profiles along the wire length for modes in the three frequency ranges during thermal transport for nanowires that are 2.4 and 4.8 *nm* long and with *M\*/M*=1 and 10. The power spectra used for Figure 3 were obtained from 200 *ps* long MD simulations after steady state is achieved. For systems with *M\*/M*=1, all modes exhibit similar temperature gradients both for *L*=2.4 *nm*, Fig. 3 (a), and 4.8 nm, Fig. 3 (b). This means that transport occurs in local thermal equilibrium with each vibrational mode having $\frac{1}{2}k_B T(z)$ of kinetic energy and confirms that thermal transport does not affect the DoS appreciably. In contrast, for *M\*/M*=10, see Fig. 3 (c) and (d), low frequency modes (that couple better with the contacts) exhibit a noticeably larger temperature gradient compared with modes of higher frequencies near the contacts. This affects a significant fraction of the 2.4 *nm* long nanowire.

Heavy contacts inject and remove low-frequency phonons, the scattering thereof leads to a temperature gradient in the corresponding low-frequency range of the nanowire spectrum. We find that for short wires scattering between low and high frequency modes is not enough for the nanowires to achieve local thermal equilibrium (even in steady state). Under these conditions the low frequency modes that couple strongly with the contacts dominate thermal transport; high frequency modes do not carry much heat and (in contrast to transport in local equilibrium) do not contribute to the effective temperature gradient either. This observation explains the first key result of our simulations, i.e., the increase in effective thermal conductivity of the wires with increasing contact mass. As the mass of the contact atoms is increased thermal transport in the wires is governed by phonons with increasingly lower frequencies that, with higher group velocities and lower scattering rates, conduct heat more efficiently than high-frequency modes. The second key result of MD simulations is somewhat expected: the effect of contact mass becomes less important as the wire length increases; the MD results quantify this expected trend. For long nanowires, scattering of phonons leads to thermal equilibration and the local temperature away from the nanowire-contact interfaces becomes frequency-independent, see Fig. 3 (d). In these cases a significant fraction of the wires is in local equilibrium and the role of contacts is less important.

**C. Interfacial resistivity**

Figure 4 (a) shows the resistivity of the upstream interface, where heat flows from the contact to the wire, and Fig. 4 (b) shows the interfacial resistivity corresponding to the downstream case. The interfacial resistivity is computed as $R = \frac{1}{J}\frac{\Delta T_{int}}{a_0}$, where $J$ is the heat flux, $a_0$ is the lattice parameter of Pt along the [112] direction, and $\Delta T_{int}$ is the temperature difference across the interface from linear extrapolation of the temperature profiles of contacting wire and substrate to the plane of interface. We find that interfacial thermal resistivity associated with the transport from wire to the contacts is size-dependent for low-frequency contacts; with short wires exhibiting lower interfacial thermal resistivity. This size dependence originates from the non-equilibrium nature of transport inside the nanowires described above. Low frequency modes in the wire can transfer heat to the downstream contact with little resistance while energy carried by high-frequency nanowire modes need to be scattered into low frequency ones before moving into the contact. Thus, short wires, where most of the heat is transferred by low frequency modes, exhibit lower contact resistance. As the wires become longer better local equilibrium is achieved within the wire and larger role of high-frequency phonons lead to an increase in interfacial resistivity. This additional resistance was denoted as *internal* in Ref. [14]. The upstream interfacial resistivity exhibits much weaker size effects.

## IV. SIMPLE MODEL ON NON-EQUILIBRIUM TRANSPORT

Based on the insight gained from MD simulations we set out to develop a simple model that captures the observed physics; in particular we would like to describe: i) how the thermal conductivity of a wire with length $L$ depends on the contact atomic mass [$\kappa_L(M^*)/\kappa_L(M)$], and ii) how this mass dependence is affected by the wire length. The effective thermal conductivity of a nanowire is defined as the ratio between the heat flux and the average temperature gradient.

Considering that during thermal transport the local temperature along the wire axis is a function of position and vibrational frequency, $T(z,\omega)$, the effective nanowire thermal conductivity can be expressed as:

$$\kappa = \frac{-J}{dT(z)/dz} = \frac{\int \kappa(\omega)\frac{dT(\omega,z)}{dz}d\omega}{\frac{d}{dz}\int \frac{C(\omega)}{\int C(\omega)d\omega}T(\omega,z)d\omega}, \quad (1)$$

where $J$ is the heat flux, $\kappa(\omega)$ is the frequency-dependent thermal conductivity that relates the temperature gradient associated with modes of frequency $\omega$ to the contribution of these modes to the total heat flux, and $C(\omega)$ is the specific heat per unit volume of the corresponding modes.

While the frequency-dependent quantities in Eq. (1) can be obtained from extensive MD simulations [4] our goal is to develop a simple and generally-applicable model. Thus we consider a linear atomic chain with harmonic interactions between first nearest neighbors for which the phonon dispersion and frequency-dependent specific heat are given by [19]:

$$\omega = 2\sqrt{\frac{K}{m}}\left|\sin\frac{a_0}{2}k\right| \quad ; \quad C(\omega) = \frac{L}{2\pi}\frac{dk}{d\omega} \quad (2)$$

where $K$ is the spring constant associated with atomic bonds of length $a_0$ and $m$ is the atomic mass ($M$ for the nanowire and $M^*$ for contacts). The wavevector $k$ in Eq. (2) spans the first Brillouin zone. Within the single mode relaxation time approximation [20], the frequency-dependent thermal conductivity can be written as a function of $k$, $\kappa(k) = C(k)v^2(k)\tau(k)$ where $C(k) = L/2\pi$ is the specific heat and the phonon group velocity in the nanowire is given by $v(k) = \frac{d\omega}{dk} = a_0\sqrt{\frac{K}{M}}\cos(\frac{a_0}{2}k)$. The phonon relaxation time $\tau$ can be considered to scale as $\omega^{-\alpha}$, which gives $\tau(k) \propto \left[2\sqrt{\dfrac{K}{M}}\sin(\dfrac{a_0}{2}k)\right]^{-\alpha}$ with exponent $\alpha$ in the range of 0 to 4 depending on the dominant scattering mechanism [21]. For thin crystalline nanowires, MD simulations show that $\alpha \sim 1$ [4]. We find that the choice of $\alpha$ does not affect the ability of the model to reproduce the MD data and, thus, we will use $\alpha = 1$.

Based on the MD results we describe the frequency-dependent temperature gradient using a function that is constant up to a characteristic cutoff frequency $\omega_c$ and monotonically decreases for higher frequencies, $\dfrac{dT(\omega,z)}{dz} = \left(\dfrac{dT}{dz}\right)_0 f(\omega)$ where

$$f(\omega) = \begin{cases} 1 & \omega \leq \omega_c(M^*) \\ \dfrac{2}{e^{[\omega(k)-\omega_c(M^*)]/[2\sqrt{\frac{K}{M}}\beta(L)]}+1} & \omega > \omega_c(M^*) \end{cases} \quad (3)$$

$\omega_c(M^*)$ is a characteristic cutoff phonon frequency and the dimensionless parameter $\beta$ controls the decay rate of the temperature gradient beyond $\omega_c$. We take the cutoff frequency as the highest vibrational frequency in the contacts, $\omega_c = 2\sqrt{\dfrac{K}{M^*}}$. We expect $\beta$ to increase (slower decay of the temperature gradient with frequency) with wire length $L$ as scattering leads to increased local equilibration.

With all the terms in Eq. (1) defined the only free parameter in the model is $\beta(L)$; this parameter is optimized to reproduce the MD data of $\kappa(M^*,L)/\kappa(M,L)$ for each wire length. The dashed lines in Fig. 2 show the results of the model together with MD data for structures with $M^*/M>1$; with a single fitting parameter our 1D model accurately describes the effect of contacts for a wide mass range. The dependence of $\beta$ on wire length obtained from fitting the MD results is shown in Fig. 5 (inset) together with the resultant frequency-dependent temperature gradient [Eq. (3)] for wires with $M^*/M=10$. As expected, $\beta$ increases monotonically with $L$ lending credence to the physics of our model. The relationship can be described using a third order polynomial $\beta(L) = L(0.00006L^2 - 0.004L + 0.11)$ with $L$ in nm, from which the behavior of wires of any length can be obtained.

For structures with $M^*/M<1$ our MD simulations show a continuing decrease in effective thermal conductivity with decreasing contact mass. A vibrational analysis shows that phonon modes of different frequencies exhibit similar temperature gradients even for the shortest wires. Consequently, additional mechanisms are contributing to the reduction in thermal conductivity of these wires, which shows a weaker dependence on the wire length; these effects are currently being investigated [22].

## V. CONCLUSIONS

In summary, MD simulations show that thermal conduction in nanoscale materials can occur away from local equilibrium and that this leads to changes in the effective thermal conductivity. For nanowires surrounded by low-frequency contacts, phonons near the interfaces are driven out of local equilibrium during thermal transport and low frequency phonons play a disproportionately dominant role in thermal transport. This leads to a significant increase in the effective nanowire thermal conductivity as the characteristic frequency of the contacts is decreased and indicates that models that assume local equilibrium are inappropriate for such cases. The downstream wire-contact interfacial conductivity also shows strong size effects; a decrease in resistivity is observed with decreasing nanowire length as transport in the wire becomes increasingly dominated by low frequency modes that can couple well to the contacts. The change in maximum phonon wavelength supported in the system may also play a role in the size effects observed as it does in superlattice systems [23]. We also developed a simple model based on phonon dispersion relationships corresponding to a linear atomic chain that captures the salient results of the NEMD simulations.

To the best of our knowledge there has been no systematic experimental investigation of the effects of contacts on the thermal conductivity of nanoscale wires and our simulations provide insight into the expected effects of interest in low-dimensional thermoelectrics and thermal management in nanoelectronics.. Our findings can also contribute to the fundamental origin of the observed size effects in phonon thermal conductivity in superlattices with respect to their periodic lengths [7,24]. As superlattice period is decreased, the number of interfaces increases, leading to a reduction in thermal conductivity [25]. However, as the thickness of each layer approaches a few nanometers our results indicate an increasingly important role of low-frequency phonons in layers of relatively light atomic mass (equivalent to short wires with heavy substrates), which increases thermal conduction within these layers. Thus, thermal conduction away from local equilibrium may play a role in the observed conductivity minimum with laminate period and the increase for lower periodicities.

Figure captions

Fig. 1 A typical nanowire-contact structure studied in this work. The Pt nanowire is oriented in the [112] direction and is epitaxially placed between two Pt* contacts.

Fig. 2 Thermal conductivity of nanowires of different lengths $L$ with contact-wire atomic mass ratio $M^*/M$. MD data are shown in symbols; model descriptions with $\alpha = 1.0$ and parameterized $\beta$ values are shown in dashed lines for structures with $M^*/M>1$.

Fig. 3 Local temperatures of modes of low, intermediate, and high frequencies along wire length during thermal conduction in steady state, with (a) $L$=2.4 $nm$, $M^*/M$=1, (b) $L$=4.8 $nm$, $M^*/M$=1, (c) $L$=2.4 $nm$, $M^*/M$=10, and (d) $L$=4.8 $nm$, $M^*/M$=10. Inset of (c) is the vibrational power spectra of contacts and nanowire in the corresponding structure. The nanowire power spectrum is divided into regimes of low (L), intermediate (I), and high (H) frequencies.

Fig. 4 Interfacial thermal resistivity of the (a) upstream interface, where heat flows from the contact to the wire, and (b) downstream interface, where heat flows from the wire to the contact.

Fig. 5 Frequency dependence of temperature gradient for wires of different lengths, with $M^*/M$=10. The inset shows $\beta$ that increases monotonically with wire length $L$.

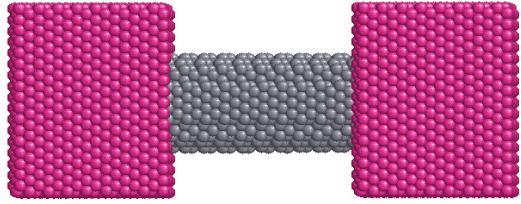

Fig. 1

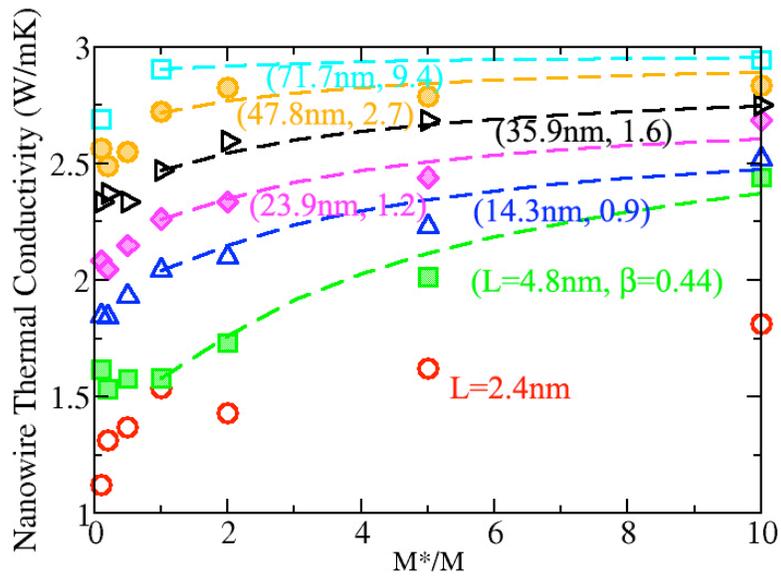

Fig. 2

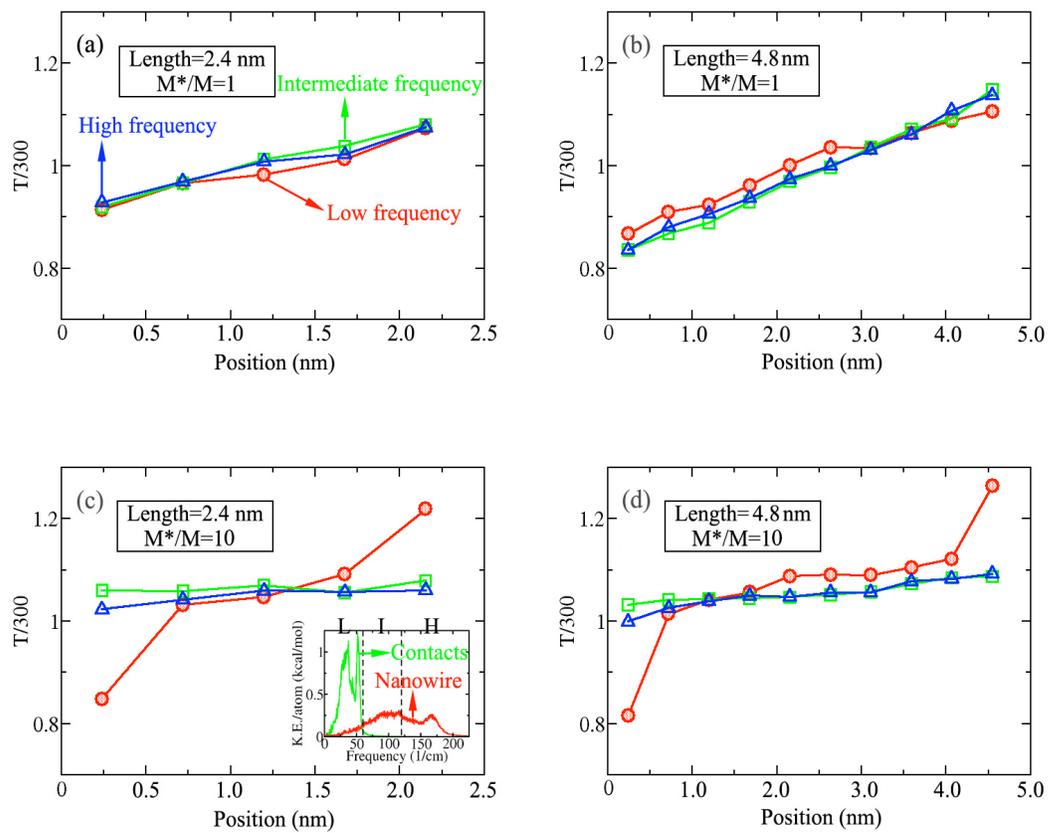

Fig. 3

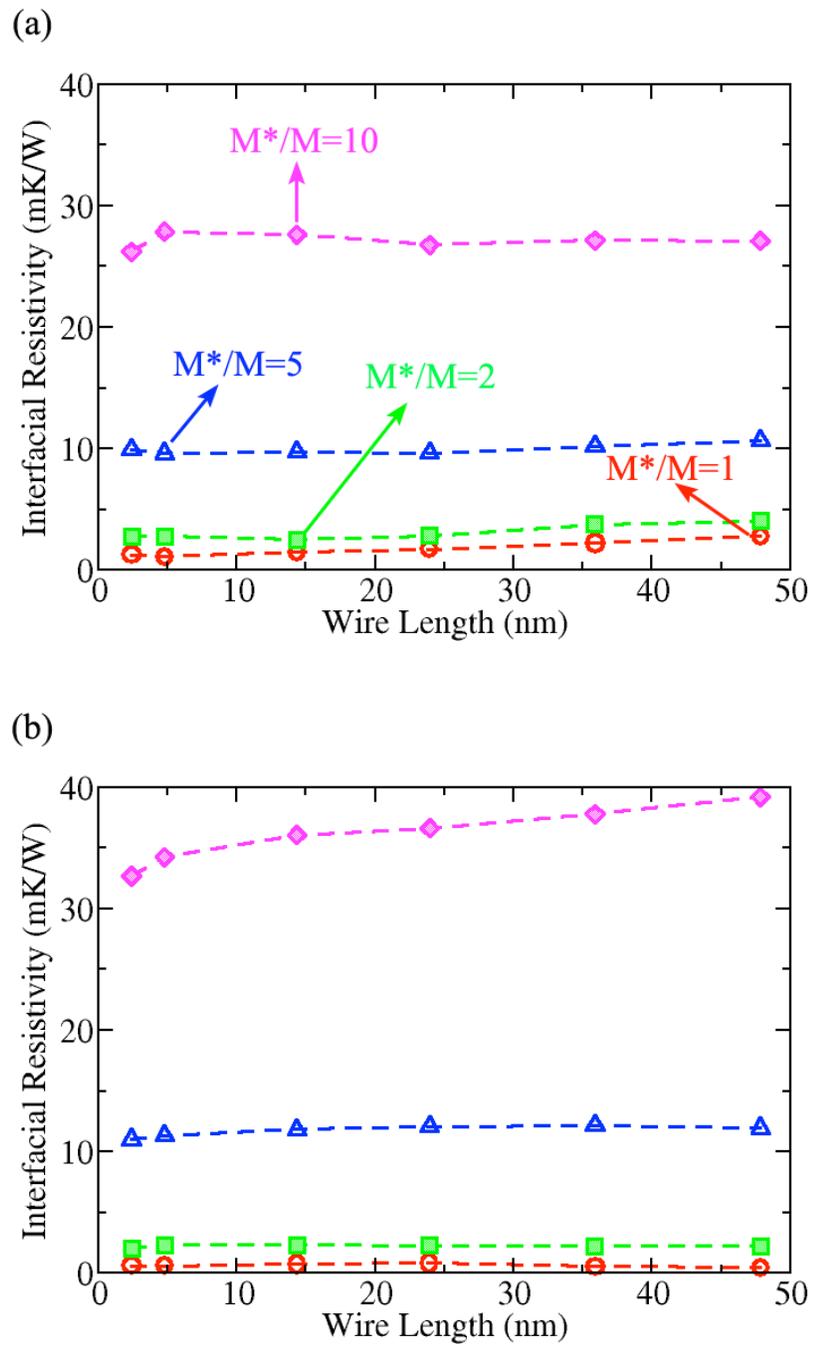

Fig. 4

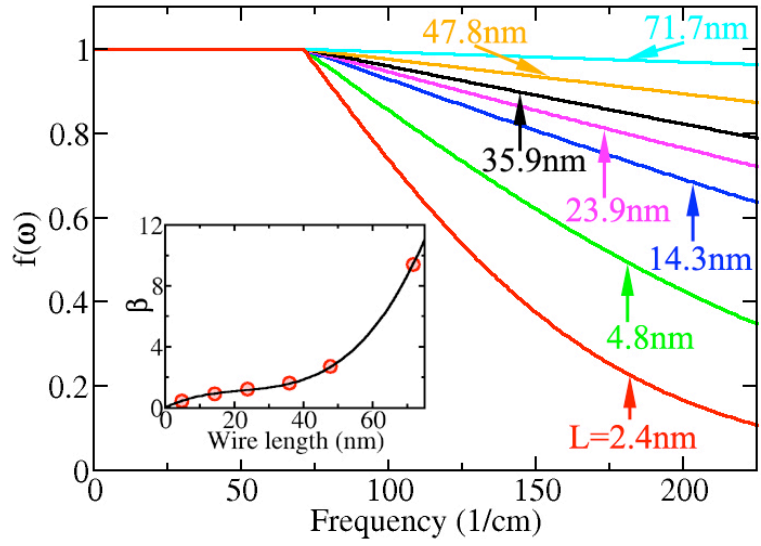

Fig. 5